\documentclass[conference]{IEEEtran}
\IEEEoverridecommandlockouts
\usepackage{cite}
\usepackage{amsmath,amssymb,amsfonts}
\usepackage{algorithmic}
\usepackage{graphicx}
\usepackage{textcomp}
\usepackage{xcolor}
\usepackage{textcomp}
\usepackage{subcaption}
\usepackage{url}
\usepackage{booktabs}
\usepackage{siunitx}

\def\BibTeX{{\rm B\kern-.05em{\sc i\kern-.025em b}\kern-.08em
    T\kern-.1667em\lower.7ex\hbox{E}\kern-.125emX}}
\begin{document}

\title{5GC-Bench: A Framework for Stress-Testing and Benchmarking 5G Core VNFs}

\author{
\IEEEauthorblockN{Ioannis Panitsas\textsuperscript{*\textsection}, 
Tolga O.~Atalay\textsuperscript{\textsection}, 
Dragoslav Stojadinovic\textsuperscript{\textsection}, 
Angelos Stavrou\textsuperscript{\textdagger\textsection}, 
Leandros Tassiulas\textsuperscript{*}}%
\thanks{\textsuperscript{1}All artifacts are available at \protect\url{https://github.com/panitsasi/5GC-Bench}}
\IEEEauthorblockA{\textsuperscript{*}Department of Electrical and Computer Engineering, Yale University, New Haven, CT, USA}
\IEEEauthorblockA{\textsuperscript{\textdagger}Department of Electrical and Computer Engineering, Virginia Tech, Blacksburg, VA, USA}
\IEEEauthorblockA{\textsuperscript{\textsection}A2 Labs, LLC, Arlington, VA, USA}
}

\maketitle

\begin{abstract}

The disaggregated, cloud-native design of the 5G Core (5GC) enables flexibility and scalability but introduces significant challenges. Control-plane procedures involve complex interactions across multiple Virtual Network Functions (VNFs), while the user plane must sustain diverse and resource-intensive traffic. Existing tools often benchmark these dimensions in isolation, rely on synthetic workloads, or lack visibility into fine-grained resource usage. This paper presents \textit{5GC-Bench}, a modular framework for stress-testing the 5GC under realistic workloads. \textit{5GC-Bench} jointly emulates signaling and service traffic, supporting both VNF profiling and end-to-end service-chain analysis. By  characterizing bottlenecks and resource demands, it provides actionable insights for capacity planning and performance optimization. We integrated \textit{5GC-Bench} with the OpenAirInterface (OAI) 5GC and deployed it on a real 5G testbed, demonstrating its ability to uncover resource constraints and expose cross-VNF dependencies under scenarios that mirror operational 5G deployments. To foster reproducibility and further research, we release publicly all the artifacts. \textsuperscript{1}

\end{abstract}
\begin{IEEEkeywords}
5G Core (5GC), Control Plane Signaling, User Plane Traffic, Stress-Testing, Benchmarking
\end{IEEEkeywords}

\section{Introduction}

The 5G Core (5GC) serves as the foundation of 5G networks, supporting connectivity, mobility, and service management through a modular architecture~\cite{3gpp23501}. Unlike the monolithic designs of previous generations, the 5GC adopts a cloud-native, Service-Based Architecture (SBA) in which network functions are decomposed into software-defined Virtual Network Functions (VNFs), decoupled from proprietary hardware and separated into control and user plane functions~\cite{survey1}. More specifically, on the control plane, the Access and Mobility Management Function (AMF) and Session Management Function (SMF) manage registration, mobility, and session establishment, while functions such as the Unified Data Management (UDM), User Data Repository (UDR), and Authentication Server Function (AUSF) provide subscription management, subscriber data storage, and authentication~\cite{3gpp23502}. The Network Repository Function (NRF) enables dynamic service discovery, serving as the central registry for inter-VNF communication. On the user plane, the User Plane Function (UPF) anchors traffic flows between devices and external Data Networks (DN), enforcing  Quality of Service (QoS) and routing policies. Together, these VNFs interact over standardized Service-Based Interfaces (SBIs) using HTTP/2, enabling flexible deployment and scaling across distributed cloud environments~\cite{3gpp29500}.

While the disaggregated architecture of the 5GC brings programmability and agility, it also introduces significant performance challenges \cite{survey2}. On the control plane, the decomposition into numerous service-based functions increases the volume of signaling traffic, as procedures such as user registration now require extensive interactions across control-plane VNFs~\cite{mukute2024ieeeaccess}. On the other side, in the user plane, the UPF must sustain increasingly heterogeneous and bandwidth-intensive workloads, ranging from video streaming to interactive applications, all of which demand strict QoS enforcement~\cite{atalay2022slices}. These challenges are tightly interconnected: bottlenecks in one part of the 5GC can cascade through long signaling chains, propagating performance degradation across other functions. This tight coupling between signaling overhead and user-plane demand highlights the urgent need for benchmarking frameworks that can uncover scalability constraints under diverse stress conditions.

\begin{figure}[t]
  \centering
  \includegraphics[width=0.92\columnwidth]{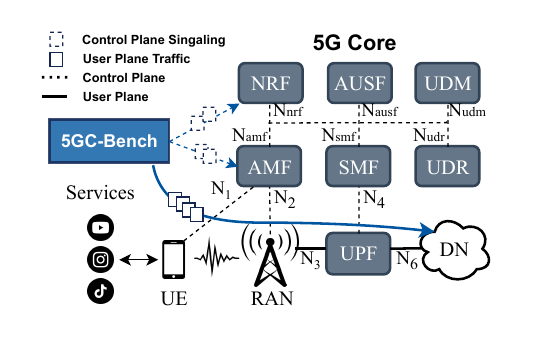} 
  \vspace{-1.2em}
  \caption{High-level view of \textit{5GC-Bench} generating signaling and traffic flows into the 5G Core for systematic stress-testing.}
  \label{fig:intro}
\end{figure}

Benchmarking and stress-testing of 5GC implementations remains challenging. Existing tools typically address isolated aspects of the system. Some focus on conformance testing and validation of signaling procedures \cite{dominato2021tutorial,mukute2024ieeeaccess, lando2023noms}, while others act as traffic generators for the user plane or emulators of the Radio Access Network (RAN) \cite{atalay2023publiccloud,atalay2022slices}. However, these approaches exhibit three limitations. First, they rarely support joint stressing of control and user plane VNFs, even though real deployments experience both dimensions concurrently. Second, they often rely on synthetic traffic patterns that fail to capture the burstiness and diversity of real-world network services. Third, they provide limited visibility into how 5GC workloads translate into resource usage across VNFs and their service-based interactions, leaving operators without actionable insights for capacity planning or bottleneck diagnosis. Without such insights, the scalability, efficiency, and resilience of 5GC deployments cannot be rigorously evaluated.

To bridge this gap, we propose \textit{5GC-Bench}, a modular benchmarking framework designed to stress-test 5GC VNFs. \textit{5GC-Bench} can be attached as an external plugin to a cloud-native core implementation, enabling comprehensive evaluation without modifications to the 5GC software stack. As illustrated in Fig.~\ref{fig:intro}, \textit{5GC-Bench} can be used to systematically generate control-plane signaling and tunnel user-plane traffic flows, supporting both synthetic workloads and trace-driven traffic profiles. Through its modular design, \textit{5GC-Bench} enables targeting individual VNFs (e.g., stressing AMF during user registration) as well as end-to-end service chains (e.g., user authentication). A telemetry subsystem continuously monitors computational, memory, and networking metrics at fine-grained resolution, aligning them with workload events to produce analysis-ready datasets for further analysis.

We integrate \textit{5GC-Bench} with the open-source 5GC implementation from OpenAirInterface (OAI)~\cite{oai} and deploy it on a real experimental 5G testbed. Our evaluation demonstrates the ability of \textit{5GC-Bench} to uncover scalability bottlenecks, highlight resource demands across control and user plane VNFs, and expose performance trade-offs under signaling-intensive and traffic-intensive scenarios. Overall, the contributions of this paper are as follows:
\begin{itemize}
    \item We propose \textit{5GC-Bench}, a modular framework for benchmarking and stress-testing of 5GC VNFs.
    \item We provide insights into the resource utilization of control and user plane VNFs under diverse workloads.  
    \item We integrate \textit{5GC-Bench} with the OAI 5GC and deploy it on an experimental 5G testbed, demonstrating its ability to systematically characterize resource demands and performance trade-offs within the core.    \item We release code, datasets, and deployment scripts\textsuperscript{1}  of \textit{5GC-Bench}  to foster reproducibility and further research. 
\end{itemize}

The remainder of this paper is organized as follows. Section II reviews related work, Section III presents the design of \textit{5GC-Bench}, Section IV outlines the experimental setup, Section V reports evaluation results, and Section VI concludes our work.

\section{Related Work}

Performance evaluation of the 5GC has received significant attention, particularly with the rise of open-source implementations \cite{oai,open5gs,free5gc}. Lando et al.\cite{lando2023noms} evaluated multiple open-source 5GC stacks, reporting  throughput and latency measurements but offering limited insights into resource consumption under complex signaling workloads. Mukute et al.\cite{mukute2024ieeeaccess} compared OAI, Open5GS~\cite{open5gs}, and free5GC~\cite{free5gc}, focusing on control-plane scalability during registration and session setup, though their scope remained narrow. Another line of research has examined 5GC deployments in cloud environments. Atalay et al.~\cite{atalay2023publiccloud} presented a performance study of the 5GC on public cloud infrastructures. Their work measured the performance of the 5GC under different deployment strategies, highlighting the impact of virtualization overheads, inter-region placement, and varying traffic loads. In a follow-up study, Atalay et al.~\cite{atalay2022slices} investigated slice scalability on a 5G testbed, analyzing how resource consumption evolves as multiple network slices are instantiated concurrently. These efforts provided valuable insights into deployment challenges and bottlenecks, but they lacked mechanisms for reproducing controlled joint control/user-plane workloads across the 5GC. Complementary studies have looked at cost-efficient evaluation aspects of the 5GC. Barbosa et al.~\cite{barbosa2024opensource} assessed open-source 5GC platforms as low-cost alternatives for academia and industry, demonstrating that they can provide a reasonable approximation of commercial cores for research purposes. Their results emphasized trade-offs between flexibility, cost, and performance, but again did not provide systematic stress-testing capabilities. Finally, Dominato et al.~\cite{dominato2021tutorial} provided a tutorial on access–core communication, clarifying signaling protocol flows but not addressing performance benchmarking concerns. Overall, prior research provides important baselines for understanding open-source 5GC implementations, and their suitability for academic experimentation. However, these works typically benchmark control and user planes in isolation, rely heavily on synthetic workloads, or offer only coarse-grained measurements. What remains missing is a unified approach that can reproduce realistic workloads while providing fine-grained visibility into the 5GC VNFs.

\section{5GC-Bench}

\subsection{System Overview}
\label{subsec:overview}

\textit{5GC-Bench} is a modular benchmarking framework for the 5GC that enables systematic stress-testing and detailed resource profiling of 5GC VNFs. With a flexible design, \textit{5GC-Bench} enables comprehensive evaluation by orchestrating control-plane signaling and tunneling user-plane flows across both isolated VNFs and full end-to-end service chains. The framework provides a declarative interface through which users define signaling procedures and traffic profiles, which are then automatically translated into control and user plane workloads. During execution, a built-in telemetry subsystem continuously captures per-VNF resource utilization and performance metrics at fine-grained temporal resolution, thereby enabling accurate and repeatable evaluation of 5GC behavior under diverse load conditions. By combining signaling procedures with realistic service traffic profiles, \textit{5GC-Bench} provides a unified instrument to: (i) quantify the scalability limits of control and user plane VNFs, (ii) localize performance bottlenecks across VNFs, and (iii) support rigorous testing and capacity planning.

\begin{figure}[t]
  \centering
  \includegraphics[width=0.9\columnwidth]{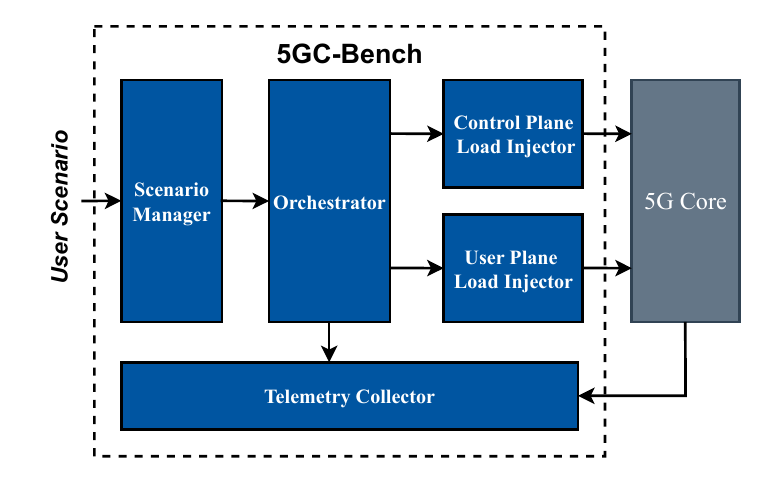} 
  \caption{High-level architecture of \textit{5GC-Bench}.}
  \label{fig:5gc-bench}
\end{figure}

\subsection{System Components}

As illustrated in Fig.~\ref{fig:5gc-bench}, \textit{5GC-Bench} is composed of five cooperating components: (i) \emph{Scenario Manager}, (ii) \emph{Orchestrator}, (iii) \emph{Control-Plane Load Injector} (CPLI), (iv) \emph{User-Plane Load Injector} (UPLI), and (v) \emph{Telemetry Collector}. These modules form a pipeline that spans the entire lifecycle of an experiment, from configuration and workload generation to execution, monitoring, and data collection. We describe the functionality of each component in the following paragraphs.

\noindent\textbf{Scenario Manager.}  
The \textit{Scenario Manager} is the entry point of \textit{5GC-Bench}, providing the user-facing interface for defining experiments. At a high level, it allows users to specify what aspects of the 5GC to stress and under which conditions. Concretely, the \textit{Scenario Manager} supports configuration of benchmark scope (control plane, user plane, or joint), selection of the 5GC VNFs under test, and choice of signaling procedures or user-plane traffic flows to emulate. It also enables fine-grained workload control, including the number of emulated base stations and connected devices, session durations, and the selection of synthetic and trace-driven traffic profiles. By capturing these options in a high-level configuration, the \textit{Scenario Manager} decouples experiment intent from execution, ensuring benchmarks are concise to define, easy to interpret, and portable across diverse 5GC deployments.

\noindent\textbf{Orchestrator.}  
The \textit{Orchestrator} acts as the control hub of \textit{5GC-Bench}, driving experiments from configuration to execution. It interprets the configuration defined in the \textit{Scenario Manager} and translates it into concrete actions on the 5GC. This includes resolving dataset dependencies, selecting control-plane procedures and user-plane traffic profiles, and initializing the relevant VNFs. The \textit{Orchestrator} also prepares the environment by instantiating emulated base stations, configuring UE contexts, and aligning workload parameters with available datasets. During execution, it coordinates the lifecycles of the \textit{CPLI} and \textit{UPLI}, enforces timing constraints, and guarantees consistent workload execution across experiments.

\noindent\textbf{Control-Plane Load Injector (CPLI).}  
The \textit{CPLI} is the dedicated module in \textit{5GC-Bench} for generating signaling load on the 5GC. It emulates control-plane interactions between devices and 5GC VNFs, driving standards-compliant procedures across the SBA of the 5GC. The \textit{CPLI} can initiate targeted procedures that stress individual VNFs (e.g., NRF discovery) or trigger end-to-end signaling chains (e.g., session setup) that traverse multiple VNFs in sequence. It supports diverse arrival patterns, including sequential requests, randomized arrivals, and bursty floods, thereby enabling controlled emulation of both typical signaling workloads and extreme overload scenarios such as large-scale signaling storms.

\noindent\textbf{User-Plane Load Injector (UPLI).}  
The \textit{UPLI} generates and tunnels user-plane workloads within the 5GC, emulating traffic exchanged between devices and the DN. Workloads can be synthetically defined by specifying flow parameters such as packet size, inter-arrival time, and session duration, or replayed from trace-driven profiles that reflect realistic service traffic. A key feature of the \textit{UPLI} is statistical multiplexing, whereby flows from multiple devices and multiple services are combined to reproduce the concurrency and diversity of traffic observed in operational networks. By leveraging open-source datasets, the \textit{UPLI} can model traffic originating from a single base station or across multiple base stations, supporting benchmarks that can capture large-scale deployment scenarios.

\noindent\textbf{Telemetry Collector.}  
The \textit{Telemetry Collector} monitors the runtime behavior of the 5GC during benchmark execution. It captures the computational, memory, and networking footprint of each VNF and the utilization of service-based interfaces at fine-grained resolution. These measurements are timestamped and aligned with control and user plane events, producing time-correlated datasets that correlate resource consumption to specific signaling procedures, and user flows across the SBA.

\section{Experimental Setup}

\begin{figure*}[ht]
    \centering
    \noindent
    \begin{minipage}{0.16\textwidth} 
        \centering
        \includegraphics[width=1\textwidth]{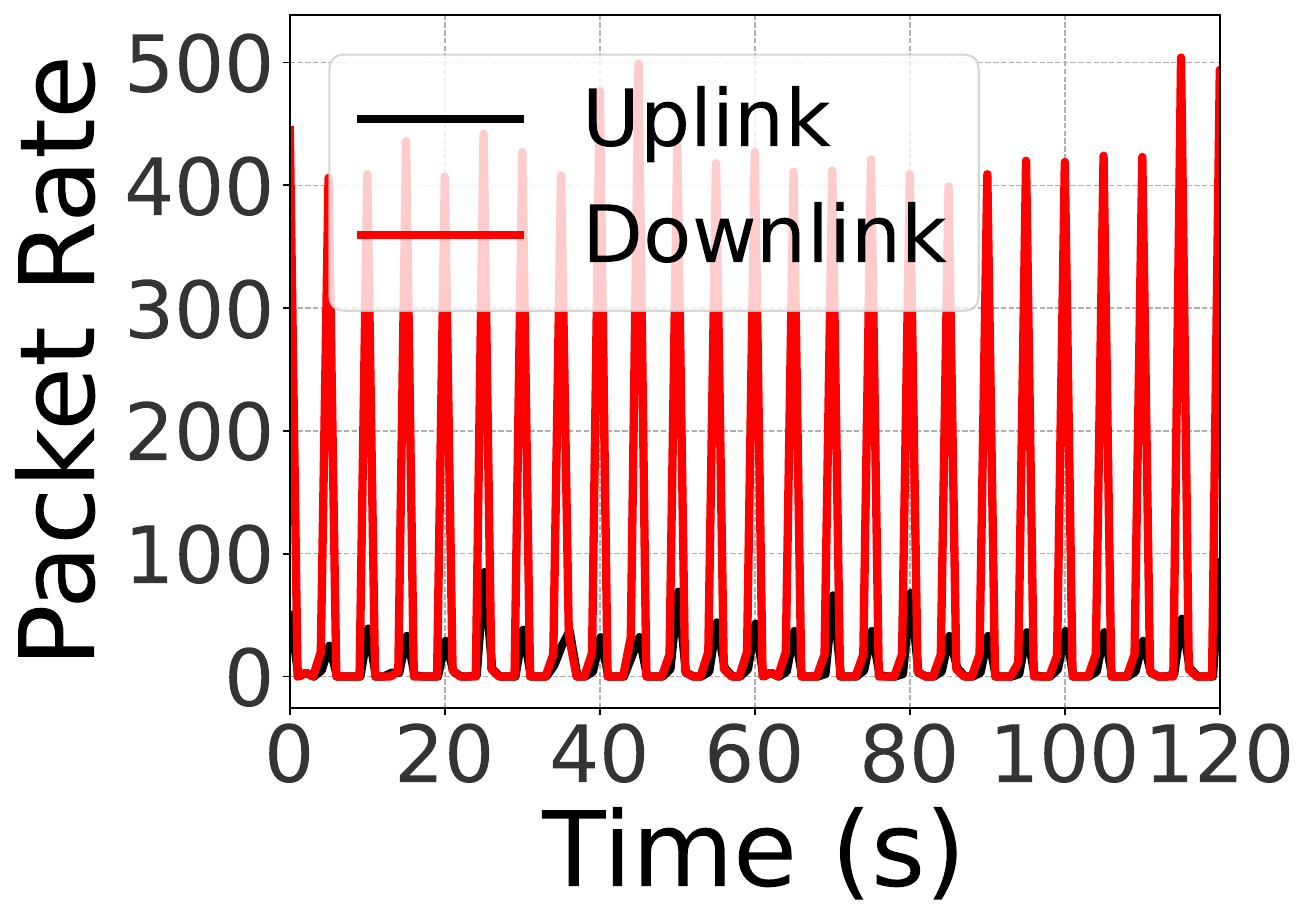} 
    \end{minipage}
    \begin{minipage}{0.16\textwidth}
        \centering
        \includegraphics[width=1\textwidth]{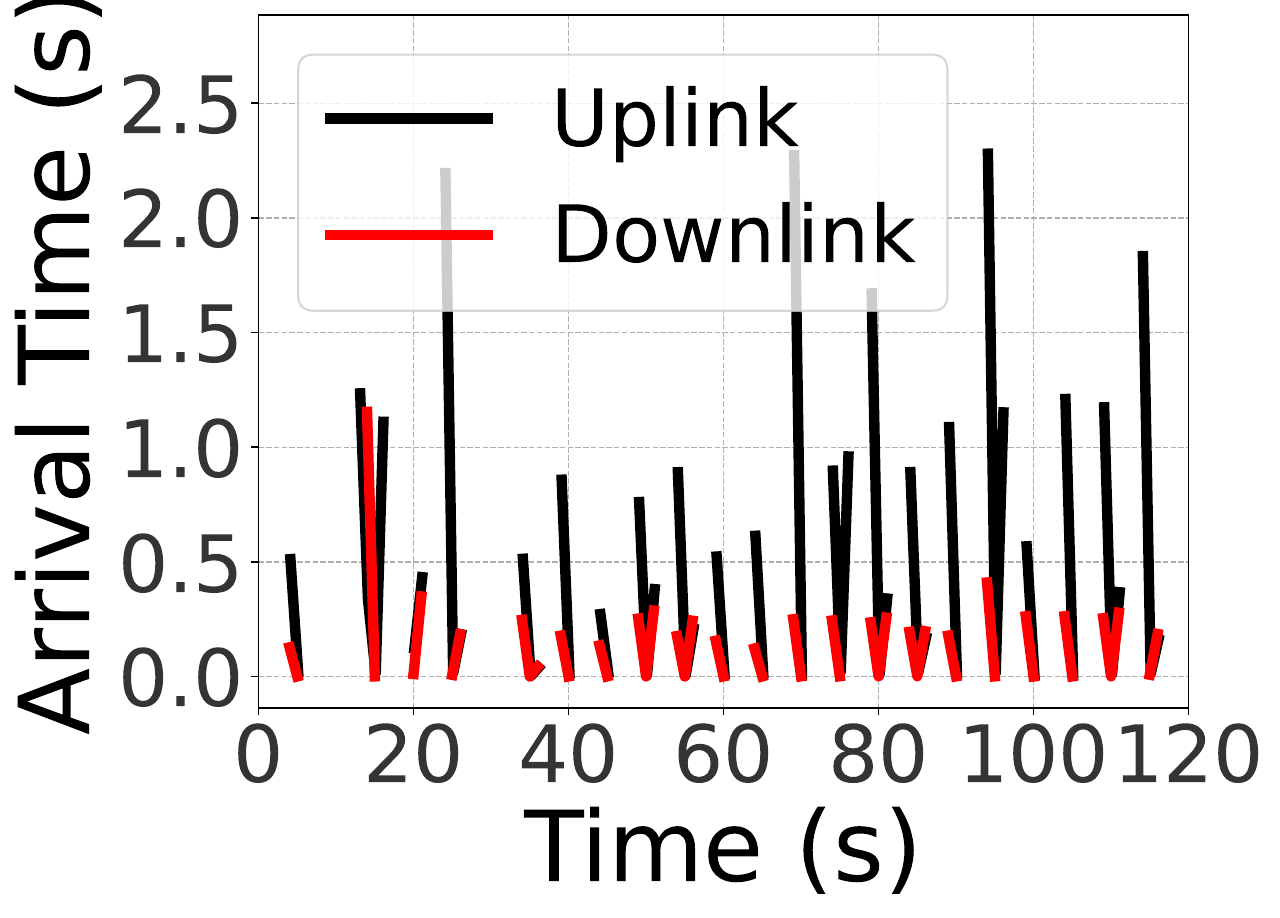}
    \end{minipage}
    \begin{minipage}{0.16\textwidth} 
        \centering
        \includegraphics[width=1\textwidth]{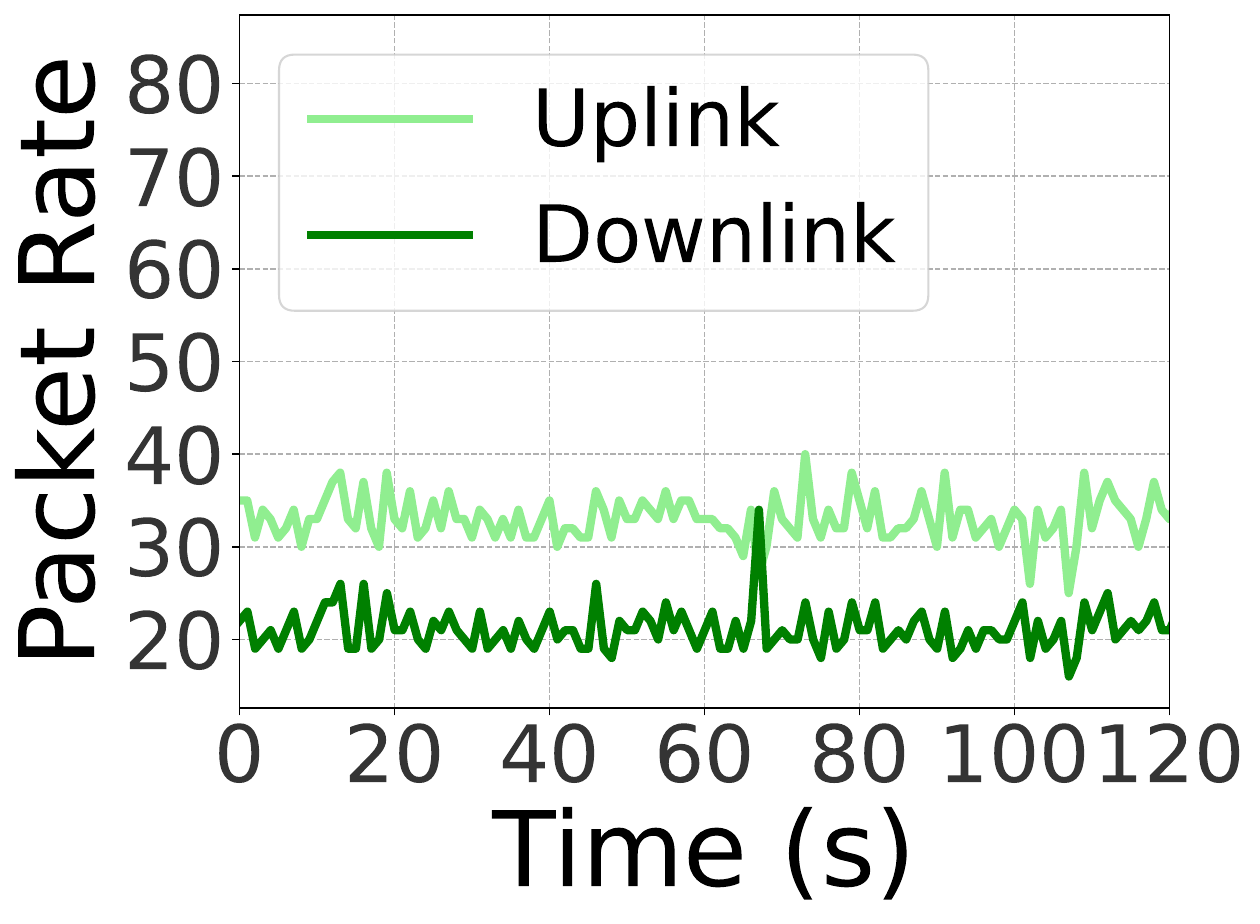}
    \end{minipage}
    \begin{minipage}{0.16\textwidth}
        \centering
        \includegraphics[width=1\textwidth]{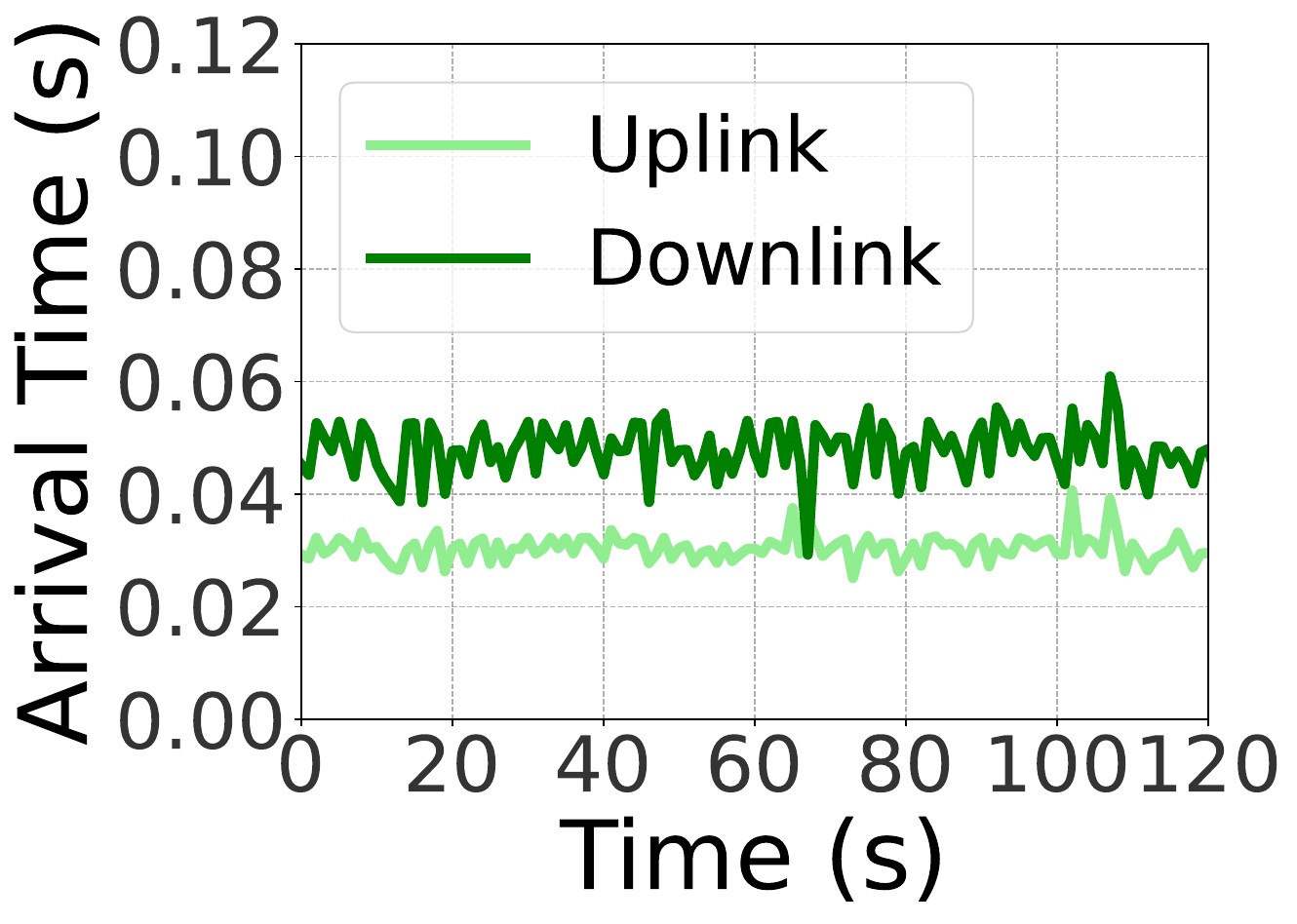}
    \end{minipage}
    \begin{minipage}{0.16\textwidth} 
        \centering
        \includegraphics[width=1\textwidth]{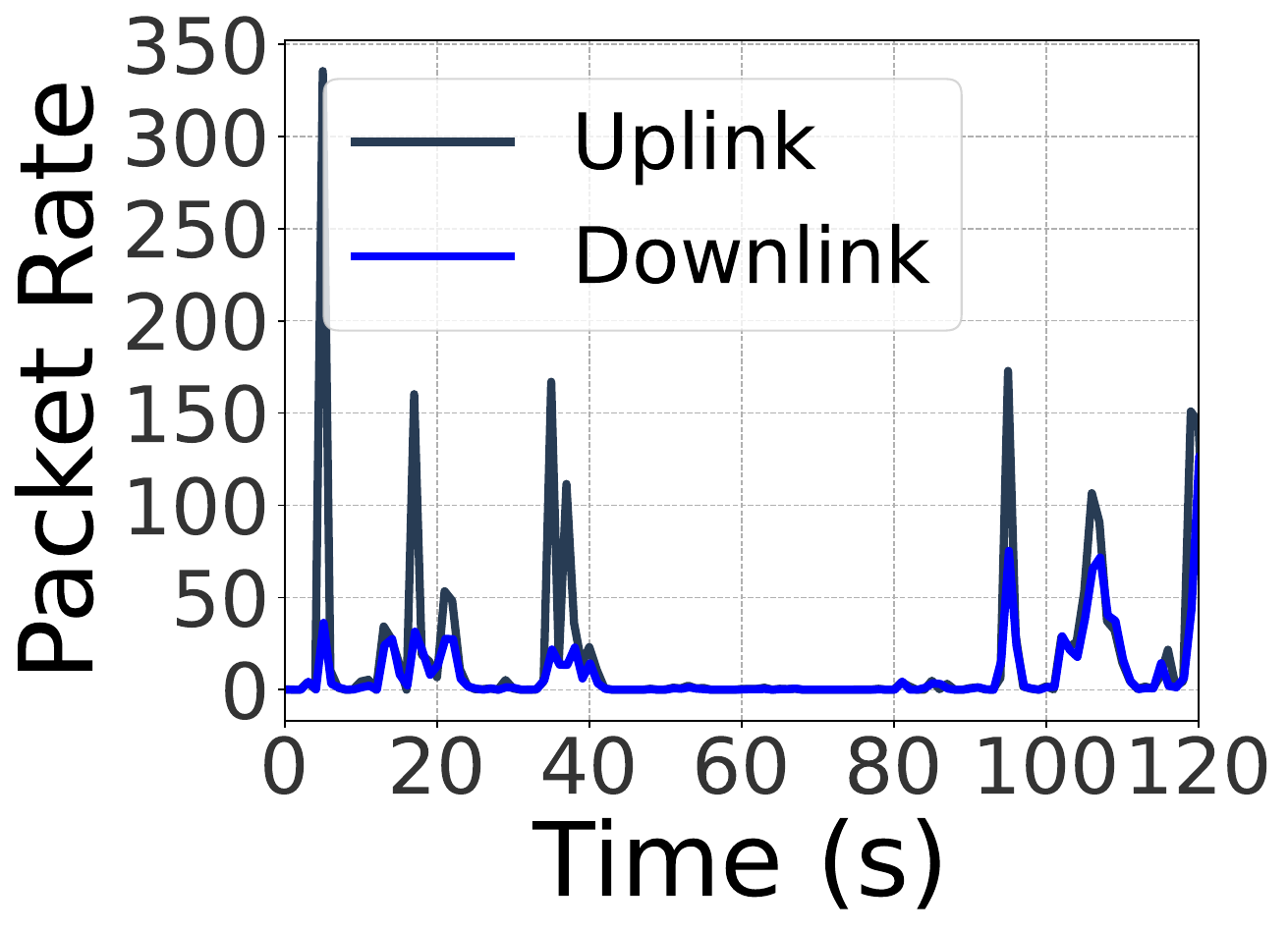}
    \end{minipage}
    \begin{minipage}{0.16\textwidth}
        \centering
        \includegraphics[width=1\textwidth]{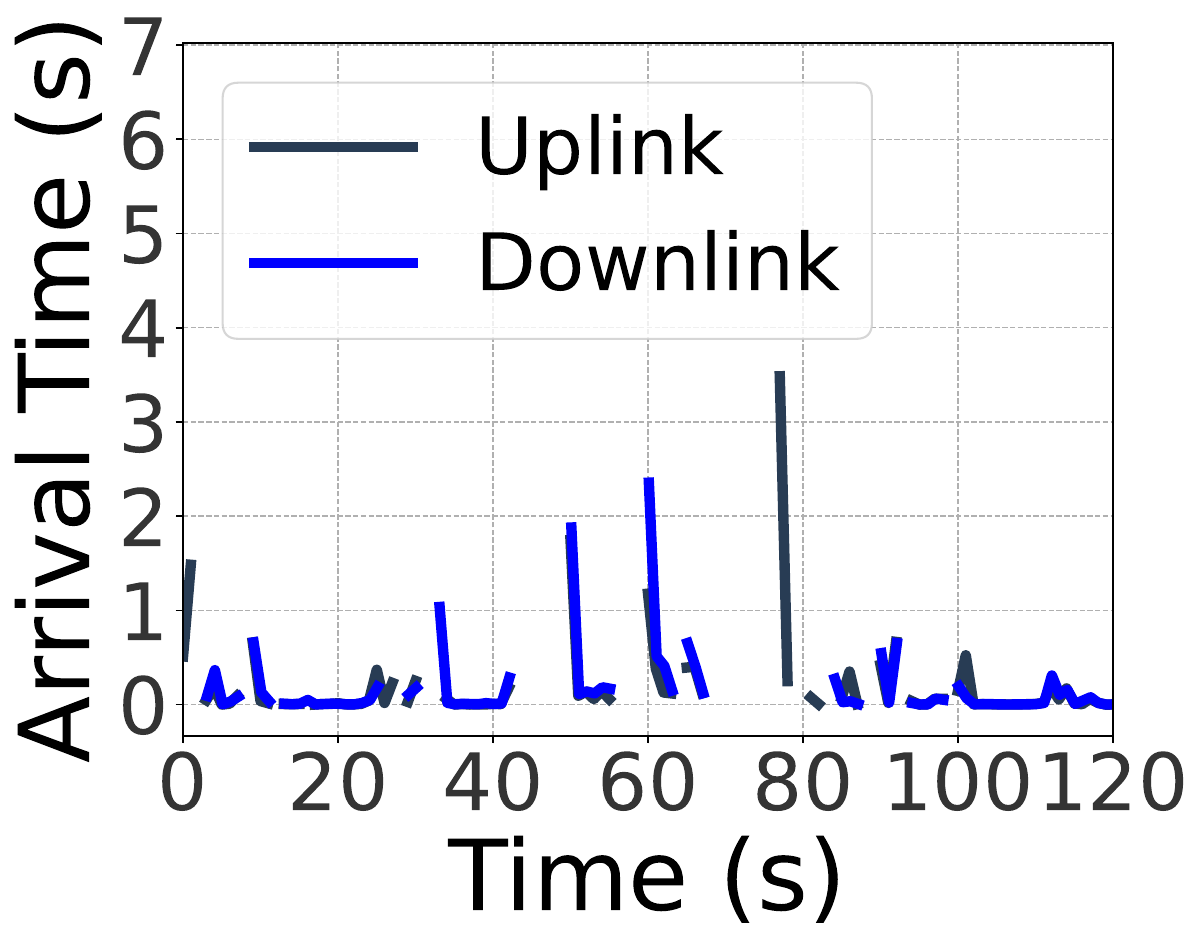}
    \end{minipage}

    \vspace{0.01cm} 
    \begin{minipage}{0.32\textwidth}
        \centering
        \text{\small (a) Video streaming (YouTube).} 
    \end{minipage}
    \begin{minipage}{0.32\textwidth}
        \centering
        \text{\small (b) Online gaming (Fortnite).} 
    \end{minipage}
    \begin{minipage}{0.32\textwidth}
        \centering
        \text{\small (c) Web browsing.} 
    \end{minipage}

    \caption{Representative service-specific traffic profiles used in our experiments.}
    \label{fig:app-traffic}

\end{figure*}

\subsection{Implementation}

We implemented \textit{5GC-Bench} in approximately $\sim$8K lines of Python and Shell code. All components are implemented in Python, except the \emph{CPLI}, which is realized in shell scripts to directly interface with the 5GC. For monitoring the resource consumption of the 5G Core VNFs, the \emph{Telemetry Collector} leverages the Docker API to provide fine-grained granularity of 1~s, recording CPU, memory, and network usage.

\subsection{5G Experimental Testbed}
\begin{figure}[t]
  \centering
  \begin{subfigure}[b]{0.4\columnwidth}
    \centering
    \includegraphics[width=\linewidth]{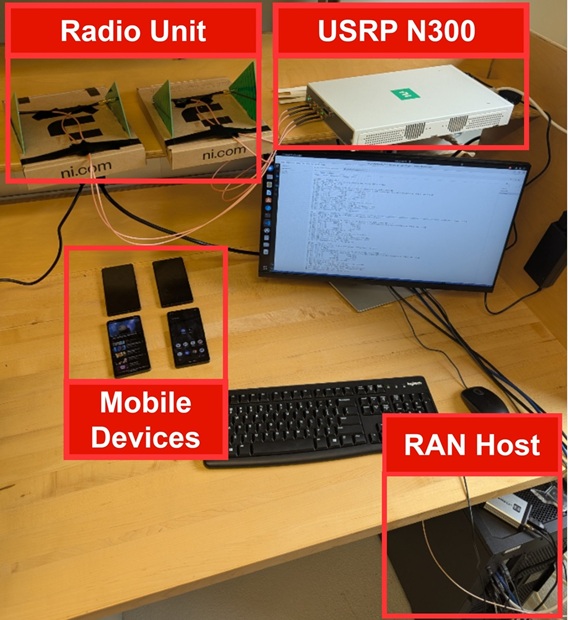}
    \label{fig:5g-testbed}
  \end{subfigure}%
  \hspace{0.02\columnwidth}
  \begin{subfigure}[b]{0.4\columnwidth}
    \centering
    \includegraphics[width=\linewidth]{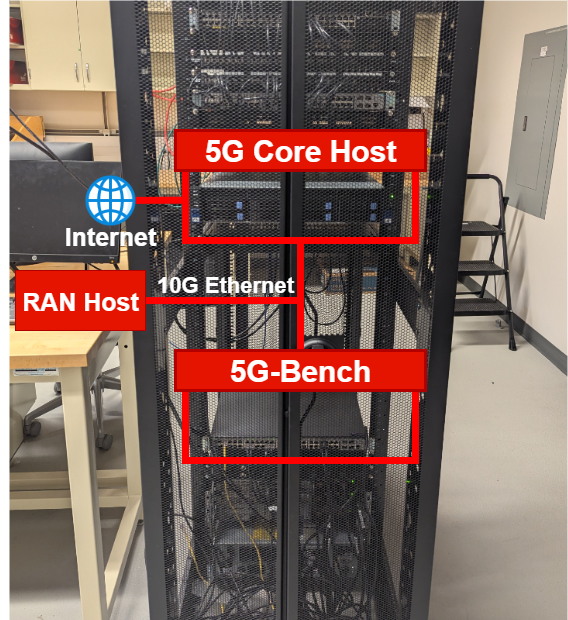}
    \label{fig:core-server}
  \end{subfigure}
  \vspace{-1em}
  \caption{5G experimental testbed.}
  \label{fig:testbed-core}
\end{figure}

We deployed \textit{5GC-Bench} on an operational 5G testbed that supports both Over-The-Air (OTA) connectivity with commercial devices and emulation with software-based devices, as illustrated in Fig.~\ref{fig:testbed-core}. The testbed is built on two commercial off-the-shelf servers hosting the core and RAN workloads, and is equipped with a software-defined radio frontend (Ettus N300) for OTA experiments. Each server features an AMD EPYC 7352 2.3~GHz 24-core processor and 128~GB of DDR4 RAM, providing sufficient compute capacity to sustain intensive workloads. Fronthaul and backhaul connections are realized using 10 Gigabit Ethernet links, ensuring high-throughput and low-latency connectivity between the RAN and the 5GC.

The 5G core was instantiated using the OAI dockerized release v2.1.0, which we selected as the reference 5GC implementation due to its standards compliance and modular containerized design. All control and user plane VNFs (AMF, SMF, NRF, AUSF, UDR, UDM, UPF) and the external DN were deployed as isolated Docker containers interconnected over a shared bridge network. On the access side, we employed \textit{gNBSIM} \cite{gnbsim} as a combined UE/RAN simulator to generate control-plane signaling traffic and tunnel user plane traffic toward the 5GC. To support realistic VNF resource evaluation, both \textit{gNBSIM} and \textit{DN} were extended with uplink and downlink traffic profiles, coupled with connection patterns derived from open-source datasets (see Section~\ref{sec:datasets}).

\subsection{Datasets}
\label{sec:datasets}

\noindent\textbf{Telecom Italia.} 
We leveraged the Telecom Italia dataset \cite{telecom_italia}, collected in Milan over 62 days, as a proxy for realistic user-plane session arrivals and control-plane signaling activity. The dataset spanned nearly 10,000 spatial cells and reported aggregated communication activity at 10-minute intervals. To prepare the dataset for use in \textit{5GC-Bench}, we merged all daily files into a single collection and aggregated values into continuous per-cell time series. We then performed a statistical analysis to classify cells into high, medium, and low load categories based on empirical percentiles of connection volumes (top 20\% as high-load, middle 60\% as medium-load, and bottom 20\% as low-load), thereby capturing different base-station utilization levels. In addition, we constructed aggregate profiles that emulated base stations serving multiple adjacent cells, thus reflecting denser deployment scenarios.

\noindent\textbf{NetMob23.}
We also employed the NetMob23 dataset \cite{netmob23}, which provided service-level mobile data traffic collected over 77 consecutive days in France. Unlike the Telecom Italia dataset, NetMob23 offered detailed information on the demand generated by 68 popular mobile services (e.g., YouTube, Netflix), aggregated at 15-minute intervals. This added the missing per-service dimension, i.e., which services (e.g., YouTube, Instagram) each connection was likely to represent. We processed the traces to derive empirical probabilities of service usage, i.e., the likelihood that a new connection was associated with a given service. These probabilities were then incorporated into \textit{5GC-Bench} to statistically multiplex user connections with the collected service traffic, thereby enabling realistic emulation of user-plane workloads.

\noindent\textbf{TelecomTS.}
Finally, we utilized TelecomTS \cite{telecomts}, a dataset collected from our OTA 5G testbed containing service traffic profiles captured from user sessions between commercial mobile phones and a base station, with each session lasting approximately 10 minutes. The dataset primarily consisted of YouTube and Twitch streaming flows and file downloads, and was further enriched with additional OTA traces from services such as Instagram, online gaming (Fortnite), and interactive browsing, enabling a more diverse mix of uplink and downlink traffic for benchmarking. Representative examples of these service-specific traffic profiles are shown in Fig.~\ref{fig:app-traffic}.

% \begin{figure}[t]
%   \centering
%   \begin{subfigure}[b]{0.48\columnwidth}
%     \centering
%     \includegraphics[width=\linewidth]{images/telecom_italia_histogram.jpg}
%     \caption{Aggregate connections.}
%   \end{subfigure}
%   \hfill
%   \begin{subfigure}[b]{0.48\columnwidth}
%     \centering
%     \includegraphics[width=\linewidth]{images/connections.jpg}
%     \caption{Daily connections in a cell.}
%   \end{subfigure}
%   \caption{Connection activity in the Telecom Italia dataset: (a) aggregate volumes across all cells and (b) temporal dynamics within a representative cell over two weeks.}
%   \label{fig:traffic-analysis}
% \end{figure}

\section{Performance Evaluation}
\textit{5GC-Bench} was employed to evaluate the 5GC along two dimensions: the control plane, where controlled signaling workloads stressed both individual VNFs and end-to-end service chains, and the user plane, where controlled service traffic was tunneled through the UPF to stress data-path processing.

\begin{figure*}[t]
  \centering
  \begin{subfigure}[b]{0.19\textwidth}
    \includegraphics[width=\linewidth]{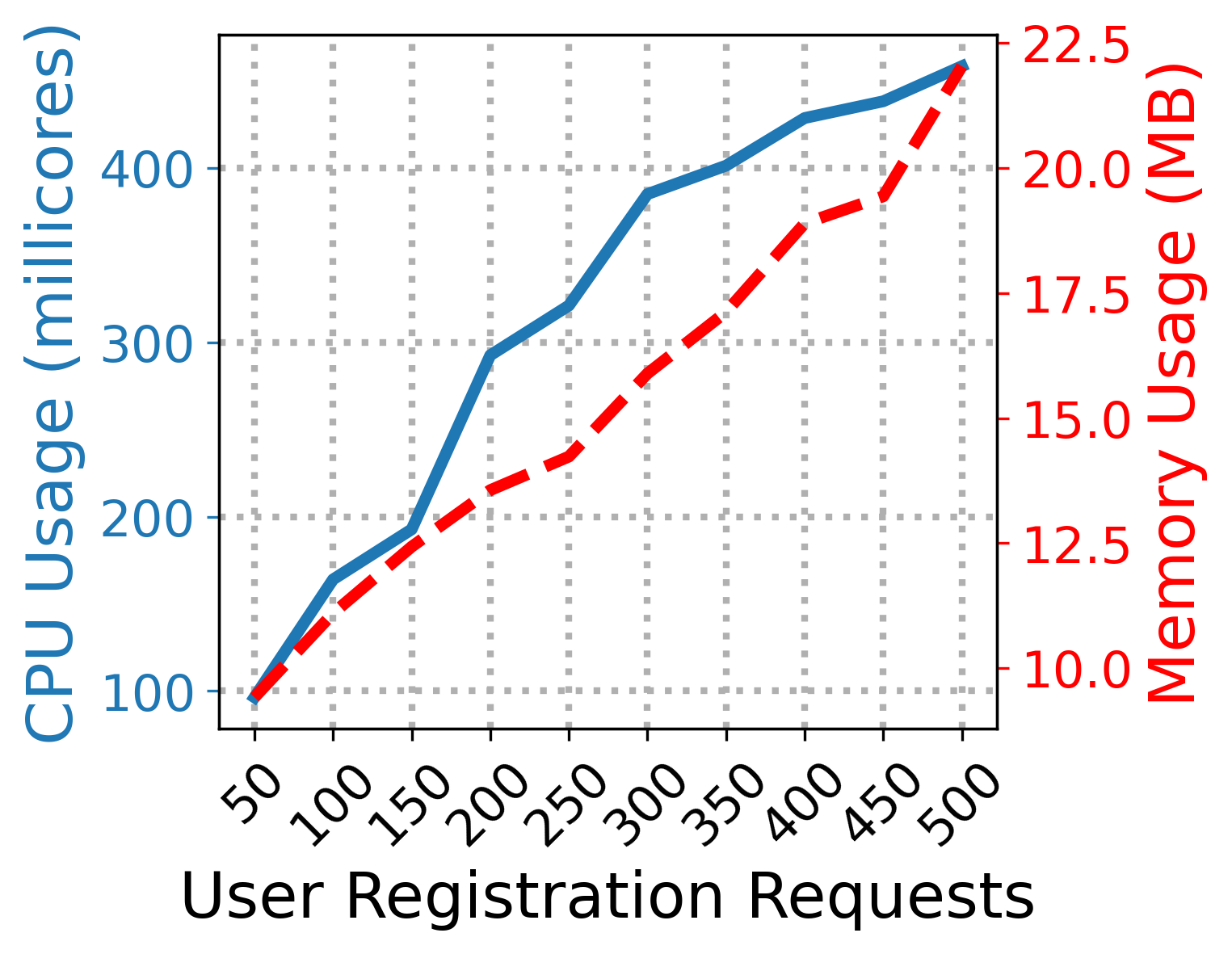}
    \caption{AMF}
    \label{fig:mem-amf}
  \end{subfigure}
  \begin{subfigure}[b]{0.19\textwidth}
    \includegraphics[width=\linewidth]{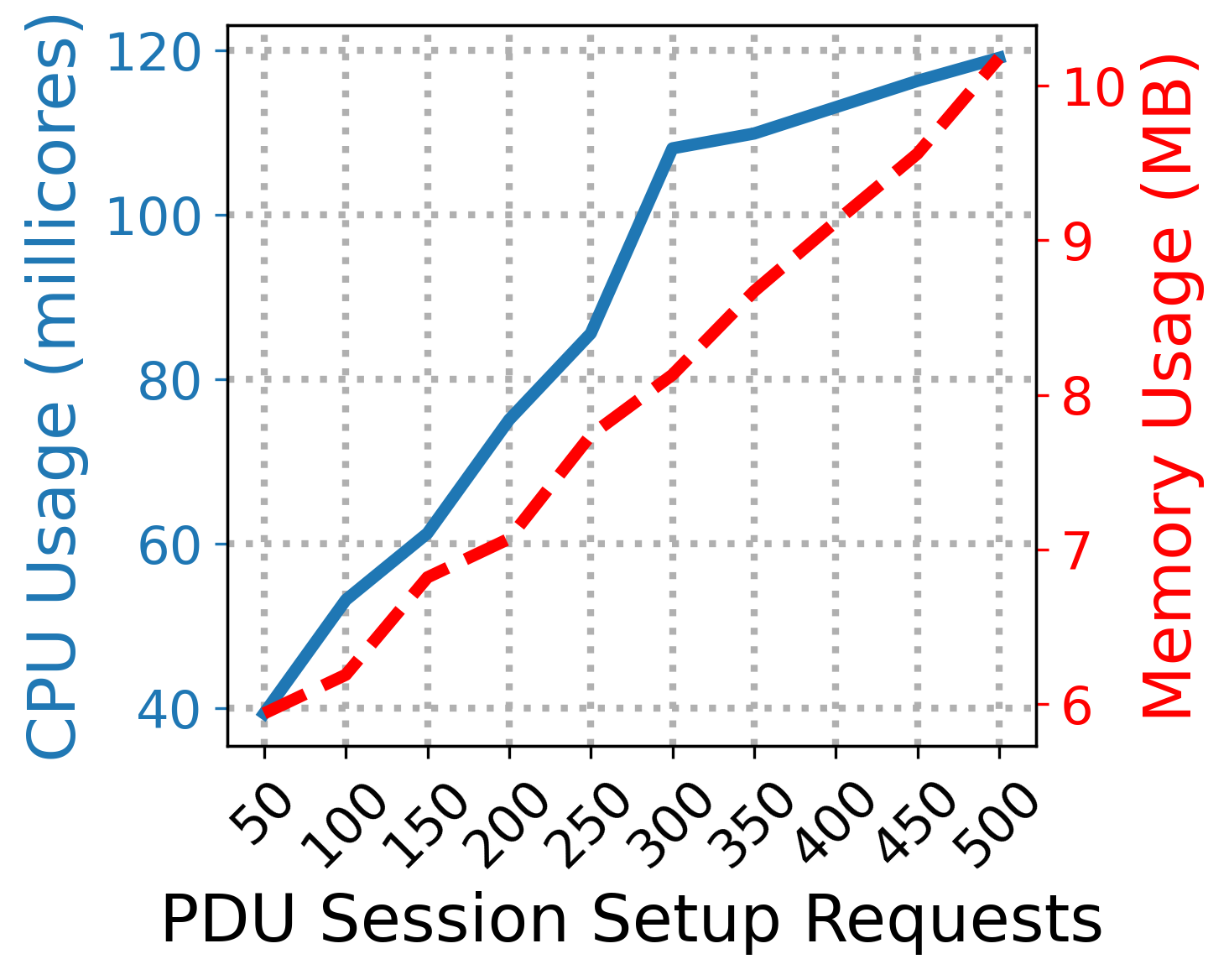}
    \caption{SMF}
    \label{fig:mem-smf}
  \end{subfigure}
  \begin{subfigure}[b]{0.19\textwidth}
    \includegraphics[width=\linewidth]{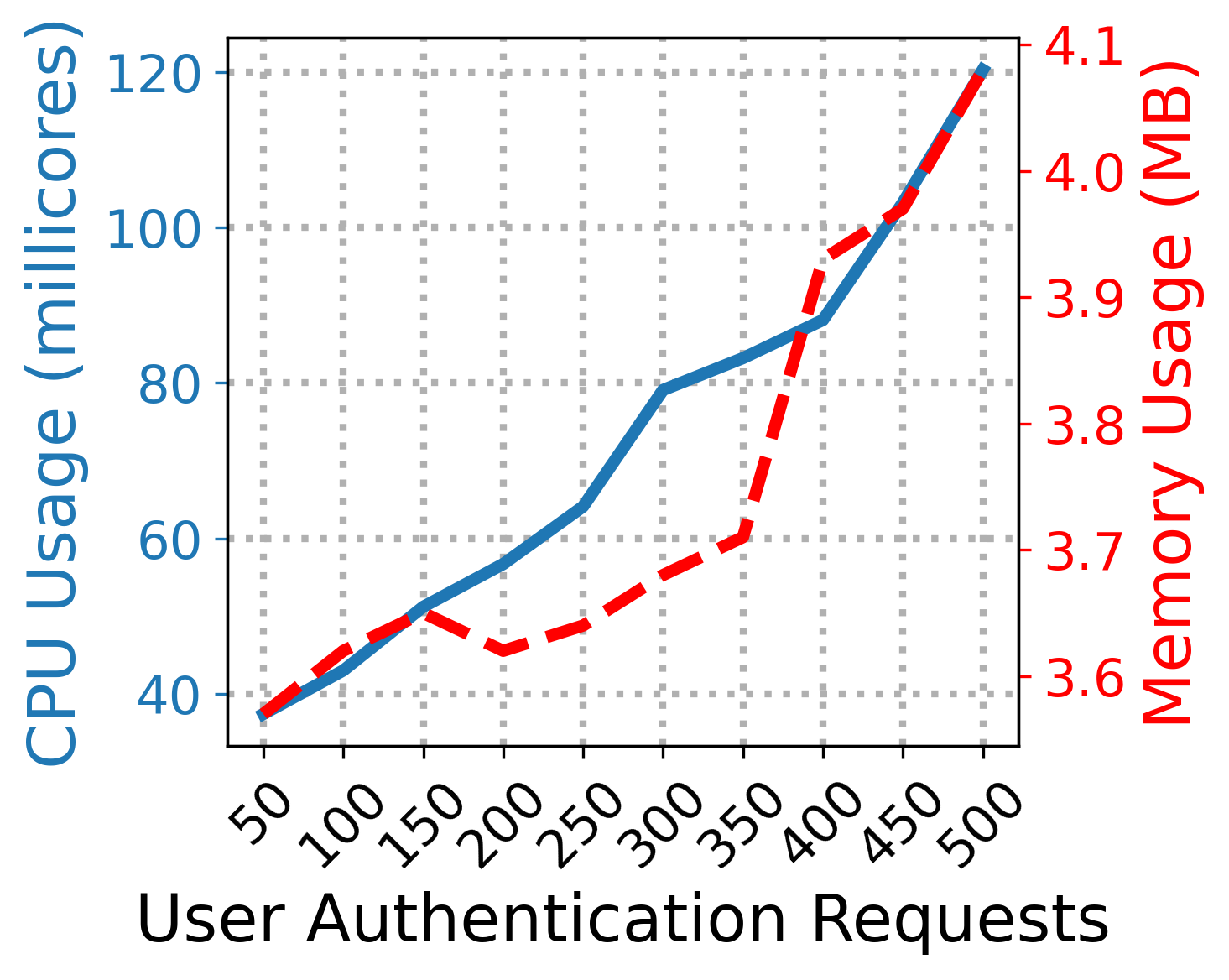}
    \caption{AUSF}
    \label{fig:mem-ausf}
  \end{subfigure}
  \begin{subfigure}[b]{0.19\textwidth}
    \includegraphics[width=\linewidth]{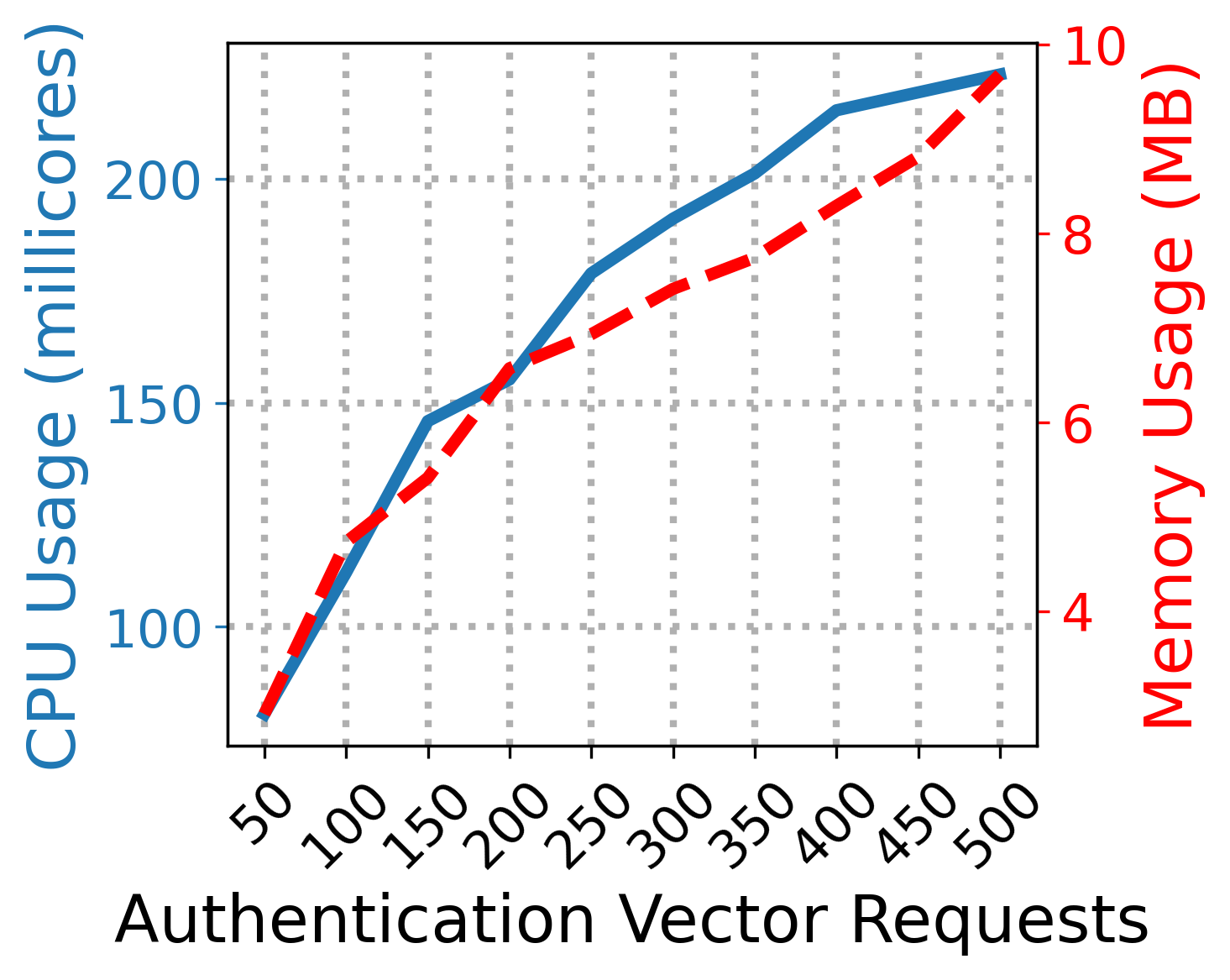}
    \caption{UDM}
    \label{fig:mem-udm}
  \end{subfigure}
  \begin{subfigure}[b]{0.19\textwidth}
    \includegraphics[width=\linewidth]{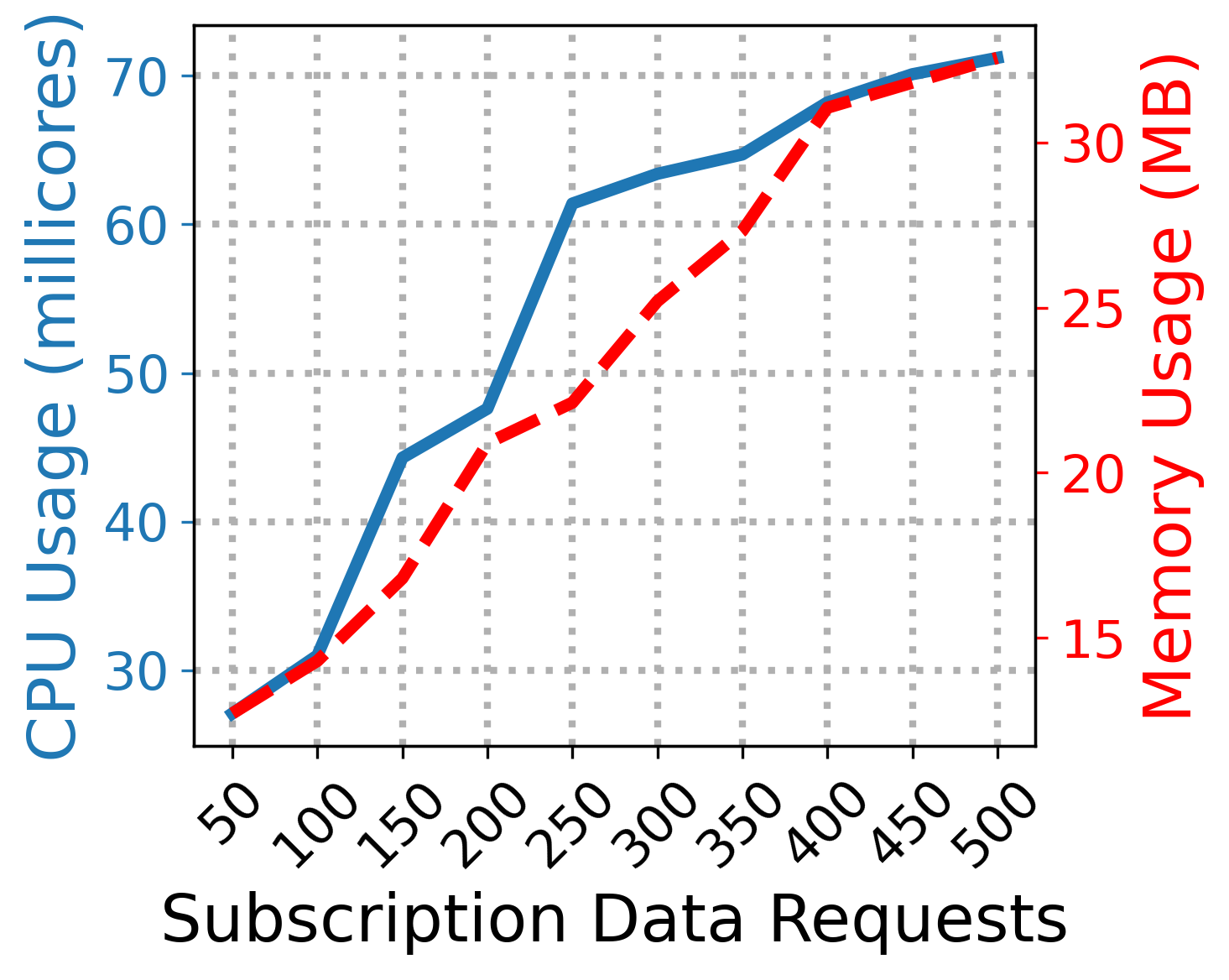}
    \caption{UDR}
    \label{fig:mem-udr}
  \end{subfigure}
  \caption{CPU and memory usage of the main 5GC VNFs under increasing request load (AMF: user registration, SMF: PDU session setup, AUSF: authentication, UDM: authentication vector generation, UDR: subscription data management).}
  \label{fig:cpu-mem-all-rows}
\end{figure*}

\subsection{Control-Plane Results}
A first set of experiments targeted the control plane VNFs to  characterize the resource demands imposed by signaling activity. For clarity, we highlight a subset of representative cases, presenting two complementary perspectives:

\noindent\textbf{Control Plane VNF Stress Analysis.}
We evaluated the performance of control-plane VNFs, examining user registration for the AMF, Packet Data Unit (PDU) session setup for the SMF, user authentication for the AUSF, authentication vector generation and subscription retrieval for the UDM, and subscription data management for the UDR. Request arrivals were derived from the Telecom Italia dataset and aggregated into fixed 10-second windows, thereby generating bursty load patterns that are representative of sudden connection surges commonly seen in cellular networks \cite{survey2}.

Figure~\ref{fig:cpu-mem-all-rows} reports CPU and memory usage of the 5GC VNFs as the signaling load increases. The results reveal clear heterogeneity in resource demands. The AMF shows a steady rise in CPU utilization with a moderate memory increase, reflecting the overhead of handling registration requests. The SMF follows a similar trend during session establishment, with CPU load scaling approximately linearly and memory usage gradually increasing as session contexts are maintained. The AUSF exhibits a consistent increase in CPU and memory usage, both driven by authentication signaling. The UDM also shows a clear upward CPU trend due to cryptographic operations for authentication vector generation, followed by a proportional rise in memory. Finally, the UDR displays moderate CPU growth but a steeper increase in memory usage, reflecting the caching and management of subscription profiles. These observations highlight that AMF, SMF, and UDM are predominantly CPU-bound, while the UDR is primarily memory-bound. Understanding these complementary patterns is critical for resource-aware placement strategies. For example, colocating CPU-intensive VNFs with memory-heavy ones may balance utilization, whereas colocating multiple CPU-bound VNFs could accelerate resource contention.

\noindent\textbf{Control Plane Service-Chain Stress Analysis.}
We also examined scenarios where control-plane signaling requires the coordinated operation of multiple VNFs, as dependencies across VNFs can amplify resource usage in ways that remain hidden under isolated profiling. A typical example is the PDU session setup, one of the most resource-intensive and latency-critical procedures in the 5GC \cite{survey2}. To emulate operational surges, we injected 200 session requests within a 10-second window, reflecting conditions that occur during mobility events or simultaneous service access at a congested base station.

\begin{figure*}[t]
  \centering
  \begin{subfigure}[b]{0.19\textwidth}
    \includegraphics[width=\linewidth]{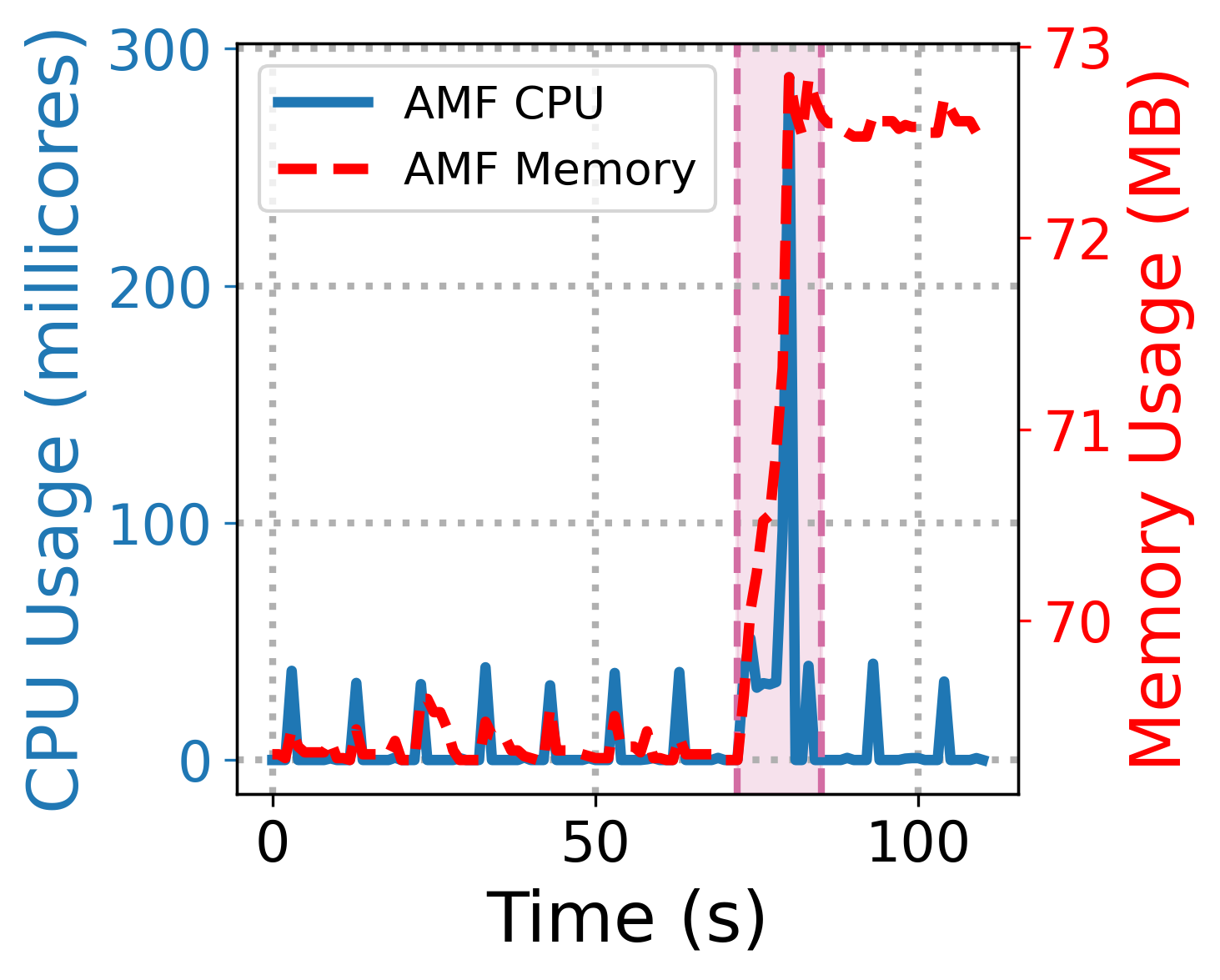}
    \caption{AMF}
    \label{fig:cpu-mem-amf}
  \end{subfigure}
  \begin{subfigure}[b]{0.19\textwidth}
    \includegraphics[width=\linewidth]{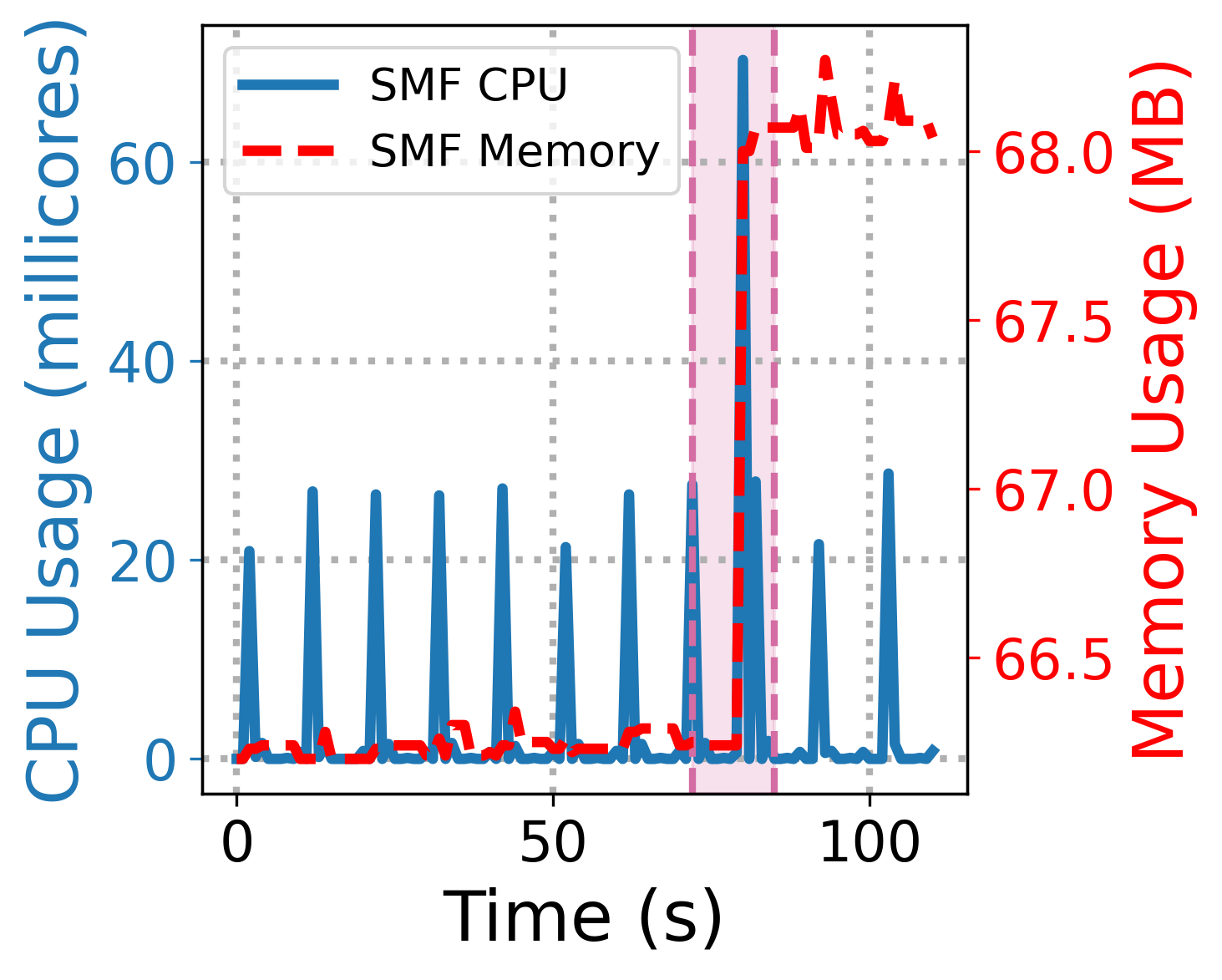}
    \caption{SMF}
    \label{fig:cpu-mem-smf}
  \end{subfigure}
  \begin{subfigure}[b]{0.19\textwidth}
    \includegraphics[width=\linewidth]{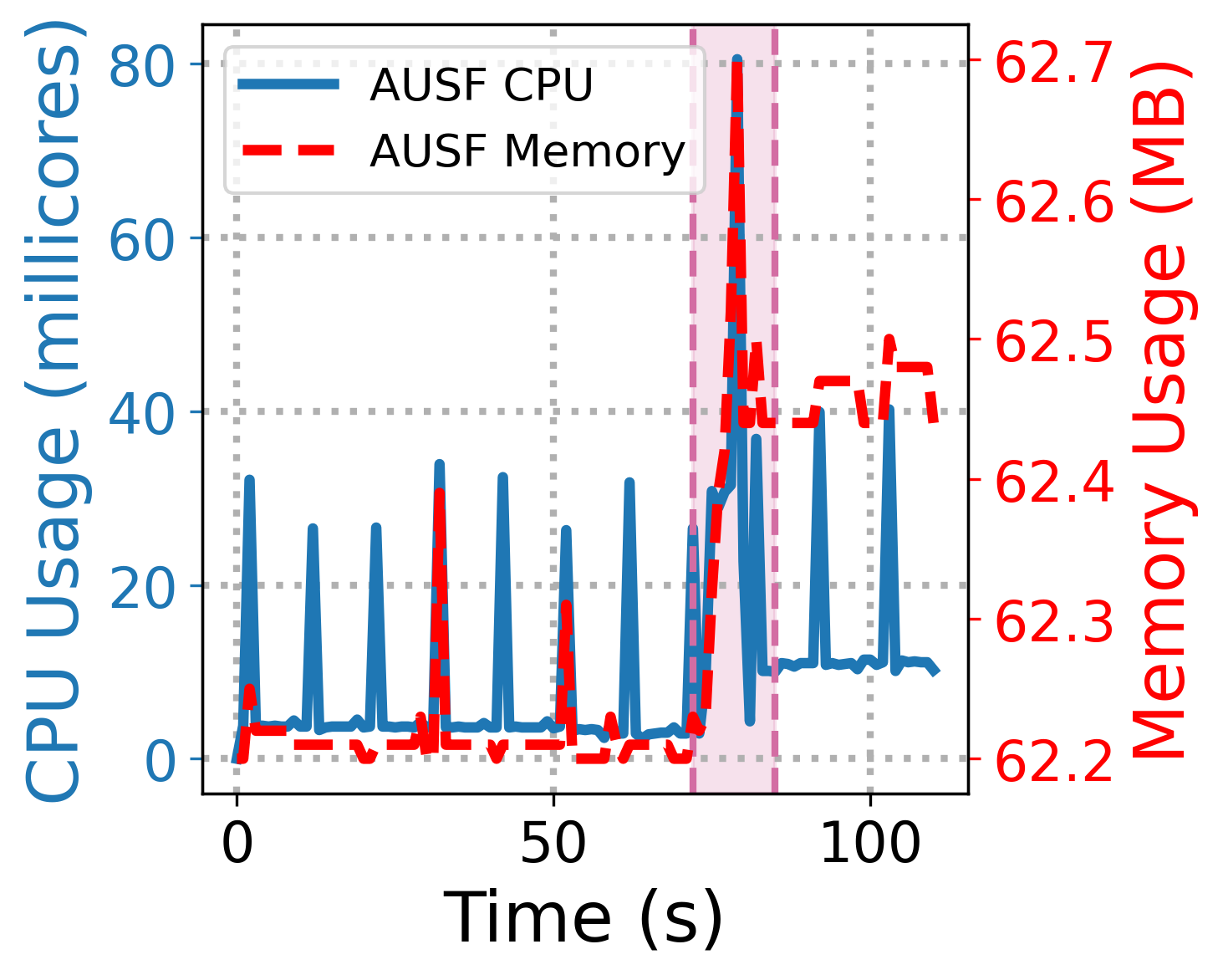}
    \caption{AUSF}
    \label{fig:cpu-mem-ausf}
  \end{subfigure}
  \begin{subfigure}[b]{0.19\textwidth}
    \includegraphics[width=\linewidth]{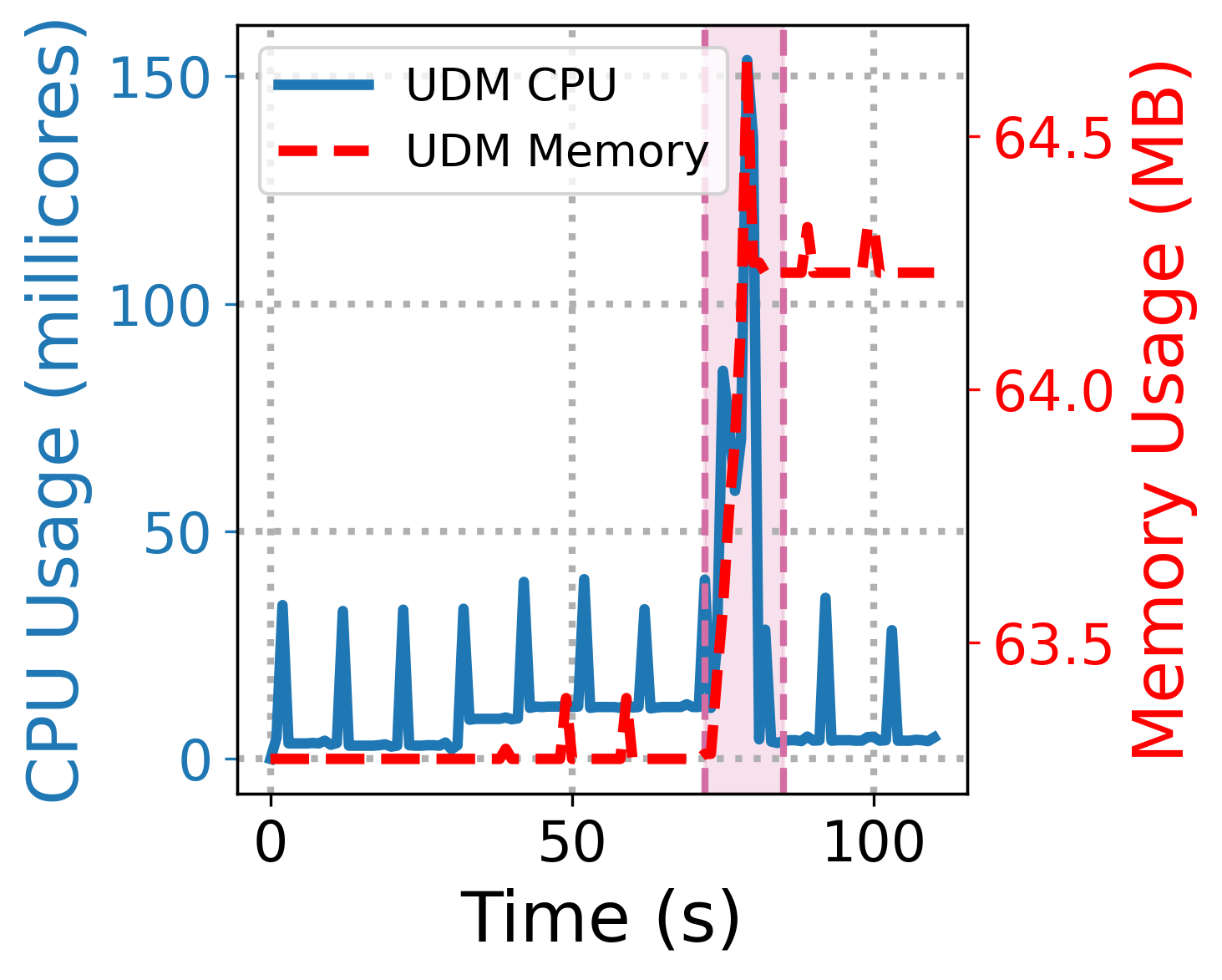}
    \caption{UDM}
    \label{fig:cpu-mem-udm}
  \end{subfigure}
  \begin{subfigure}[b]{0.19\textwidth}
    \includegraphics[width=\linewidth]{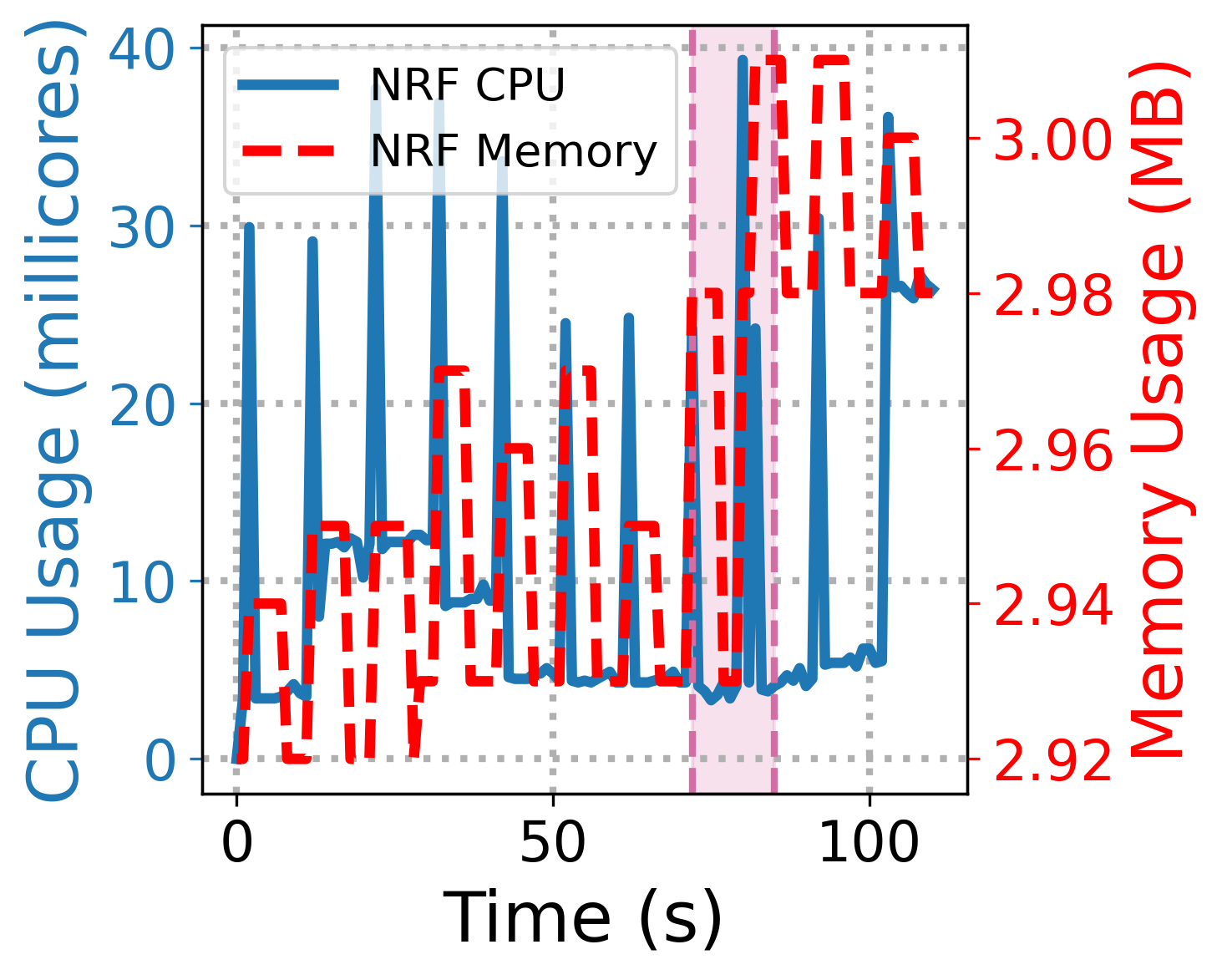}
    \caption{NRF}
    \label{fig:cpu-mem-nrf}
  \end{subfigure}
  \caption{Service-chain analysis of the PDU session setup. CPU and memory usage are shown over time, with the shaded pink region marking the time window during which bursty session establishment requests arrive for processing in the 5GC.}
  \label{fig:inter-vnf-results}
\end{figure*}

Figure~\ref{fig:inter-vnf-results} reports CPU and memory usage across the AMF, SMF, AUSF, UDM, and NRF during the session-burst interval. Recurring CPU spikes appear across all VNFs because heartbeat messages periodically probe liveness and trigger lightweight processing. When the burst of 200 session requests arrives, highlighted by the pink region, CPU utilization rises sharply across all network functions. The AMF and UDM incur the heaviest processing load, as they handle session context establishment and authentication/vector generation, respectively. The SMF contributes additional processing for session setup, the NRF experiences increased load from service discovery, and the AUSF from authentication signaling. Memory usage also rises during this interval, particularly at the AMF and SMF where temporary states are stored. After the burst, both CPU and memory stabilize at levels higher than before, reflecting the persistence of session contexts. 

\subsection{User-Plane Results}

To complement our control-plane analysis, we extended our study to the user plane and examined the computational burden of service traffic on the UPF. Leveraging the TelecomTS dataset, we replayed traffic profiles for YouTube, Instagram, browsing, and online gaming, as well as a composite workload integrating all four. We selected these profiles as prior measurement studies~\cite{netmob23} have shown they constitute the dominant application classes in operational 5G networks, providing a realistic foundation for evaluating UPF scalability.

\begin{table}[t]
\centering
\caption{UPF CPU and memory utilization across services under increasing number of user sessions.}
\label{tab:upf-split}
\scriptsize
\setlength{\tabcolsep}{13pt} 
\begin{tabular}{lccccc}
\toprule
 & \multicolumn{5}{c}{\textbf{User Sessions}} \\
\cmidrule(lr){2-6}
\textbf{Service} & \textbf{100} & \textbf{200} & \textbf{300} & \textbf{400} & \textbf{500} \\
\midrule
\multicolumn{6}{c}{\textbf{CPU Usage (millicores)}} \\
\midrule
Gaming    & 42  & 63  & 102 & 147 & 181 \\
Browsing  & 142 & 248 & 361 & 476 & 582 \\
YouTube   & 156 & 285 & 408 & 520 & 613 \\
Instagram & 324 & 477 & 656 & 801 & 916 \\
Mixed     & 225 & 404 & 554 & 663 & 752 \\
\midrule
\multicolumn{6}{c}{\textbf{Memory Usage (MB)}} \\
\midrule
Gaming    & 4.33 & 4.33 & 4.33 & 4.33 & 4.33 \\
Browsing  & 4.35 & 4.35 & 4.36 & 4.34 & 4.34 \\
YouTube   & 4.33 & 4.34 & 4.34 & 4.34 & 4.34 \\
Instagram & 4.34 & 4.35 & 4.35 & 4.35 & 4.36 \\
Mixed     & 4.35 & 4.36 & 4.36 & 4.36 & 4.36 \\
\bottomrule
\end{tabular}
\end{table}

\noindent\textbf{User Plane VNF Stress Analysis.}
We benchmarked the UPF by replaying traffic profiles for each service in isolation, while varying the number of concurrent user sessions from 100 to 500 over a 10-minute interval. Table~\ref{tab:upf-split} summarizes the resulting average CPU and memory utilization. The results show that CPU usage scales proportionally with the number of active sessions, thereby confirming that packet-processing load is the primary bottleneck in the user plane. Video services such as YouTube and Instagram impose the heaviest demand, as their sustained high-rate flows require continuous processing. Web browsing produces intermediate utilization, since the UPF must handle irregular packet bursts triggered by user interactions. Online gaming traffic remains lightweight, with small packet sizes and limited packet rates consistently translating into minimal processing requirements. The mixed workload, formed by multiplexing flows from all four applications, yields CPU consumption that reflects the aggregate traffic mix, with video sessions overwhelmingly dominating overall demand. Across all cases, memory utilization stays nearly constant at 4--5~MB, clearly indicating that UPF performance is constrained by CPU cycles rather than memory capacity. These findings confirm that the UPF is CPU-bound under realistic 5G traffic profiles and that application type directly determines the rate at which resources are consumed.

\begin{figure}[t]
  \centering
  \begin{subfigure}[b]{0.48\columnwidth}
    \centering
    \includegraphics[width=\linewidth]{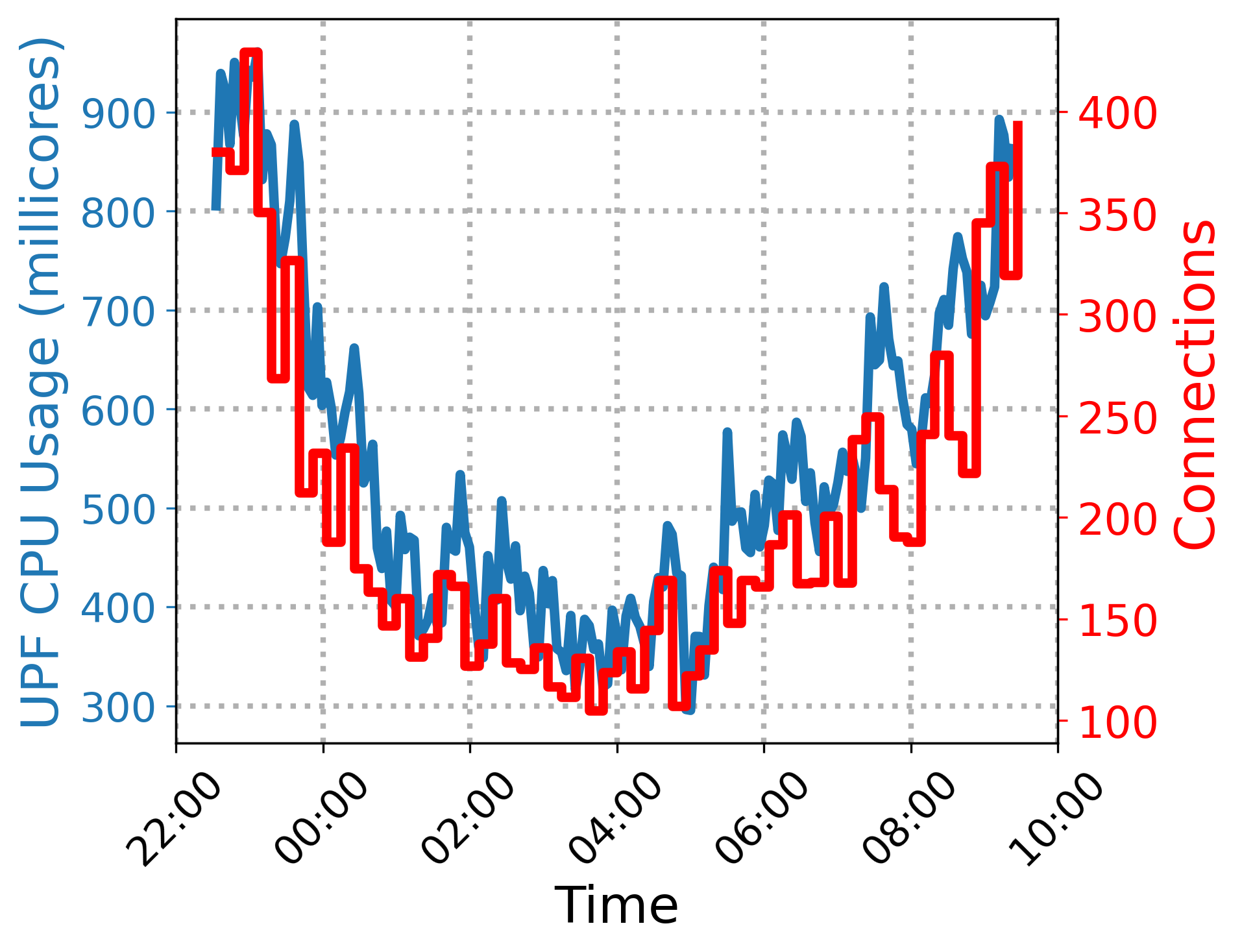}
    \caption{Low-load scenario}
    \label{fig:upf-low}
  \end{subfigure}
  \hfill
  \begin{subfigure}[b]{0.48\columnwidth}
    \centering
    \includegraphics[width=\linewidth]{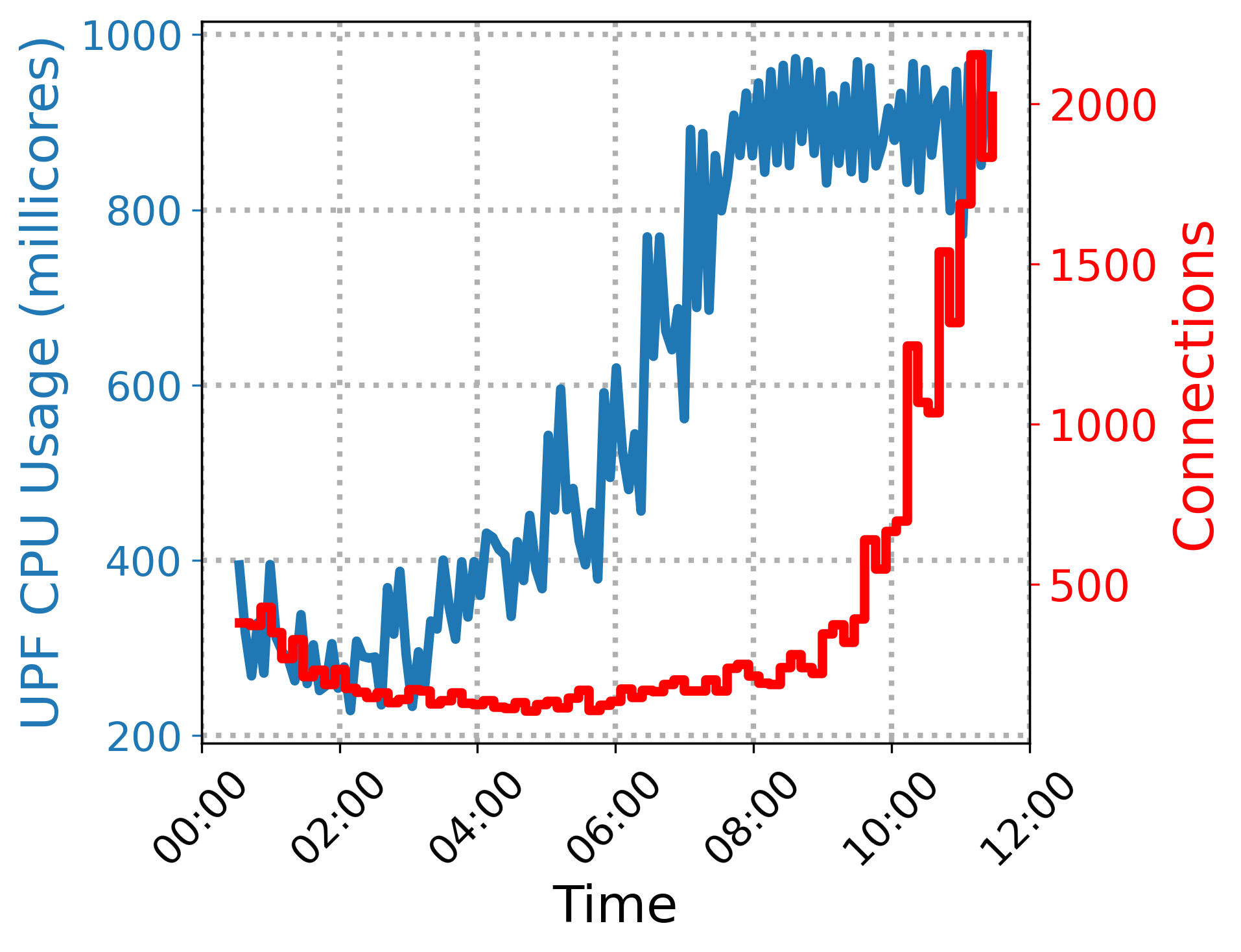}
    \caption{High-load scenario}
    \label{fig:upf-high}
  \end{subfigure}
  \caption{UPF CPU utilization over time for two representative base-station load scenarios.}
  \label{fig:upf-utilization}
\end{figure}

\noindent\textbf{UPF Profiling under Realistic Traffic Conditions.}
To evaluate UPF behavior under realistic load conditions, we considered two scenarios derived from the Telecom Italia dataset. The first corresponds to a single-cell trace, emulating the activity of one base station under low-load utilization. The second aggregates traffic from five neighboring cells, representing a denser deployment with substantially higher load. To reflect realistic service composition, traffic flows in both cases were multiplexed according to application distributions reported by the NetMob dataset. Figure~\ref{fig:upf-utilization} reports UPF CPU utilization and number of active connections over a 12-hour period. In the low-load case (Fig.\ref{fig:upf-utilization}a), CPU usage remains moderate during the night (200–400 millicores) with fewer than 400 active connections. In contrast, the high-load case (Fig.\ref{fig:upf-utilization}b) sustains substantially higher utilization, rising steadily throughout the day to exceed 800 millicores, while connections scale to nearly 2{,}000 during busy-hour activity. This combination of per-session dynamics with per-service profiles highlights how \textit{5GC-Bench} can reproduce operational traffic conditions and reveal how variations in base-station load directly map to both UPF resource consumption and connection density.

\section{Conclusion}
This paper introduced \textit{5GC-Bench}, a modular framework for stress-testing 5GC VNFs under realistic control and user plane workloads derived from open datasets and collected traces. By systematically characterizing resource demands across critical core functions, \textit{5GC-Bench} provides a practical tool for capacity planning, bottleneck diagnosis, and performance optimization. We integrated the framework with the OAI 5GC and validated it on a real 5G testbed, demonstrating its effectiveness for comprehensive core network evaluation.

\bibliographystyle{IEEEtran}  
\bibliography{references}          

\end{document}